\documentclass[12 pt, amssymb,prb,showpacs]{revtex4}
\usepackage{epsfig}
\usepackage{dcolumn}
\usepackage{amsmath}
\begin{document}
\title{\bf\ Beating pattern in radiation-induced oscillatory magnetoresistance in 2DES: coupling of plasmon-like and
 acoustic phonon modes.}
\author{J. I\~narrea$^{1,2}$}
\affiliation{$^1$Escuela Polit\'ecnica
Superior,Universidad Carlos III,Leganes,Madrid,28911,Spain\\
$^2$Unidad Asociada al Instituto de Ciencia de Materiales, CSIC,
Cantoblanco,Madrid,28049,Spain.}
\date{\today}
\begin{abstract}
We present a microscopic theory  on the observation of a beating pattern in the
radiation-induced magnetoresistance oscillations at very low magnetic field. We consider that such a beating pattern
develops as a result of the coupling between two oscillatory components:
the first is a system of electron Landau states
being harmonically driven by radiation. The second is a lattice oscillation, i.e.,
 an acoustic phonon mode. We analyze the dependence of the
beating pattern on temperature, radiation frequency and power.
We conclude that the beating pattern is an evidence of the radiation-driven
nature of the irradiated Landau states that makes them behave as a collective plasma oscillation
at the radiation frequency. Thus, the frequency of such plasmons could be tuned from
microwave to terahertz in the same nanodevice with an apparent technological application.

\end{abstract}
\maketitle
Beating patterns show up when there are two oscillatory contributions coexisting
and coupled in the same physical system.
For instance, beating patterns can be observed in magnetoresistance ($R_{xx}$)  of two-dimensional
electron systems (2DES) when there are two populated conduction electron subbands involved in the
transport. This situation gives rise to the well-known  magneto-intersubband
scattering oscillations (MISO)\cite{miso1,miso2}.
 They can also be obtained in $R_{xx}$ of 2DES
with strong Rashba spin-orbit coupling\cite{rashba1,rashba2}. For both cases the two oscillatory subsystems are the two sets of broadened Landau levels
of slightly different energies.
On the other hand, two important physical  effects in magnetotransport of 2DES were
discovered more than a decade ago by Mani et al.,\cite{mani1}: the radiation-induced magnetoresistance oscillations
(RIRO) and the even more
striking of zero resistance states (ZRS)\cite{mani1,zudov1}.
\begin{figure}
\centering\epsfxsize=2.8in \epsfysize=3.2in
\epsffile{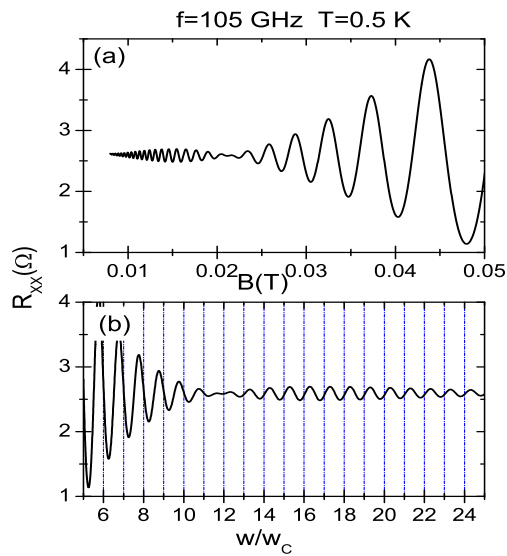}
\caption{ Calculated $R_{xx}$ for a radiation frequency of 105 GHz. In panel a) $R_{xx}$
vs $B$ and in b) $R_{xx}$ vs $w/w_{c}$. Apart from the usual RIRO we observe a beat at low $B$
with a node around $B=0.02T$ in panel a). In panel b) we observe a change in phase difference of
$\pi$ when crossing the node.}
\end{figure}
To date, there is not a clear  consensus  on the physical origin
of such remarkable effects. After
a huge number of experiments \cite{mani01,mani02,mani2,
mani3,smet,mani03,mani04,mani05,mani5,wiedmann1,wiedmann2,mani6,mani61,mani62,ye,ganichev1,ganichev2}
and  proposed theories\cite{ryzhii,girvin,lei,rivera,vavilov,ryz1,ryz2,ryz3,ina2,ina21,ina22,ina23,ina24,ina4,ina41,ina42,ina5,ina51,ina71,ina72,plat1,plat2,plat3,plat4,veltukov}
to explain them, we have to admit that they are still under debate.

Soon after the discovery of RIRO and ZRS, another surprising experimental result regarding RIRO  was published\cite{mani100}.
It consisted in an unexpected beating pattern at very low $B$ superimposed to RIRO.
As with RIRO this beating pattern was radiation-induced.
This subtle effect was
overlooked by the scientific community
 and very little attention was paid.
Nevertheless, there has recently  been shown new experimental evidence presenting
 a similar beating pattern profile on RIRO at very low $B$\cite{shi2} too.
As explained above, this radiation-induced beating patter indicates the presence
of two comparable oscillatory contributions. However, possible physical scenarios giving rise
to beating patterns, such as two populated electron
subbands or Rashba spin-orbit coupling,  cannot easily explain the obtained
experimental results\cite{mani100,shi2}.

In this letter  we develop a microscopic theory to explain the beating pattern in RIRO based, on
the one hand,  on
the physical effect of plasmon-phonon coupling in polar semiconductors\cite{varga,kunie,hase}.
According to it, in polar semiconductors like GaAs,
collective oscillations of electrical charges (plasmons) and lattice ions oscillations (phonons) can
couple via Coulomb interaction. As a result, the initially individual (plasmon and phonon) modes
re-adjust their frequencies to give rise to new hybrid plasmon-phonon coupled modes.
On the other hand, our microscopic theory is based on the previous model for RIRO,
 {\it the radiation-driven electron orbit model}.
 This model, in turn,  is based on
the exact solution of the electronic wave function in the presence
of a static magnetic field  and radiation.
In this model the electrons orbits or  Landau States (LS)
move back and forth harmonically    driven by  radiation, (driven-LS), at the radiation frequency ($w$).
Thus, the guiding centers of the LS perform harmonic and {\it classical} trajectories making
 the system of driven-LS behave like a collective oscillation of electric charge, i.e.
a plasmon-like mode.
Now, this plasmon-like mode, with acoustic frequency, can couple with a collective
lattice ions oscillation of similar amplitude and frequency, an acoustic phonon mode.
Thus, we can observe the rise of a beating pattern,
 for instance, in  $R_{xx}$. Therefore, the observation
of beats in the RIRO profile would be a clear evidence of the spatial swinging nature
of the irradiated LS. This provides a source of excitation of acoustic plasmon-like
modes in 2DES with a frequency ranging from the microwave (MW) to the terahertz (THz) part of the
spectrum.

\begin{figure}
\centering \epsfxsize=2.8in \epsfysize=3.in
\epsffile{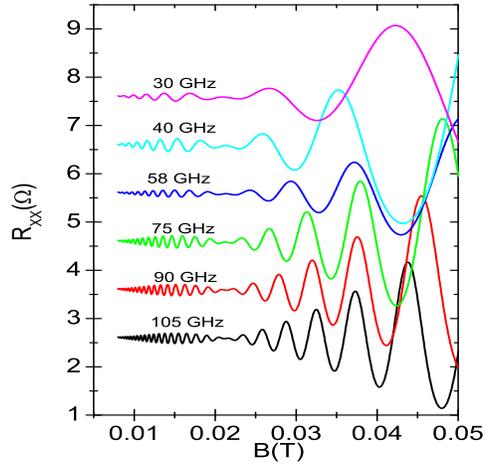}
\caption{Dependence of the beating pattern
on the radiation frequency ranging from 30 GHz to 105 GHz. The node
position does not change irrespective of the radiation frequency.}
\end{figure}
\begin{figure}
\centering \epsfxsize=2.8in \epsfysize=3.0in
\epsffile{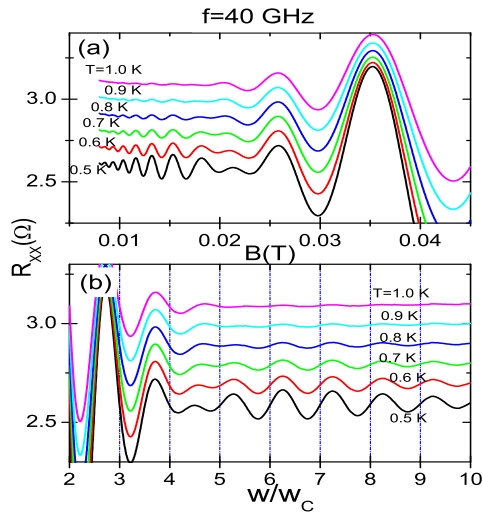}
\caption{Dependence of the beating pattern on temperature. In the upper panel
we exhibit irradiated $R_{xx}$ vs $B$ and in the lower panel the same vs $w/w_{c}$. In the lower
panel the node position moves to lower $w/w_{c}$ for decreasing $T$. In the upper panel we observe
the opposite trend, the node moves to higher $B$ for decreasing $T$. }
\end{figure}

As we said above, {\it the radiation-driven electron orbits model}\cite{ina2,ina21} was developed to explain
the striking effects of RIRO and ZRS. One of the main conclusion of this theory is that
under radiation the LS oscillate, with their guiding centers, at the radiation frequency according to
$X(t)=X_{0}+A \sin wt$, where $X(t)$ is the time dependent LS guiding center position,
$X_{0}$ is the same without radiation and
$A=\frac{e E_{o}}{m^{*}\sqrt{(w_{c}^{2}-w^{2})^{2}+\gamma^{4}}}$
where, in turn, $E_{0}$ is the  radiation electric field and
$w_{c}$ the cyclotron frequency.
$\gamma$ is a phenomenologically introduced damping factor
for the electron scattering  with the lattice ions.
Following the physics of plasmon-phonon coupling
we consider  that the system of driven-LS, behaving like a "plasmon-like" mode,  can couple with an acoustic phonon mode
of similar frequency and amplitude.
We derive classically and similarly to a
system of coupled harmonic oscillators, the new frequencies  of the
hybrid modes\cite{novotny} and the guiding center position of the driven-LS mode. Firstly, and after some algebra,
these frequencies are given by\cite{olson,lee,torma}:
\begin{equation}
2w_{\pm}^{2}=(w^{2}+w_{ac}^{2})\pm \sqrt{(w^{2}-w_{ac}^{2})+16 \lambda^{2}w_{ac}w}
\end{equation}
where $w$ is the  frequency of both, radiation and the driven-LS mode. $w_{ac}$ is the frequency of the acoustic phonon mode
and $\lambda$ the plasmon-phonon coupling constant.
If $w$ is only slightly different from $w_{ac}$, i.e., $w\simeq w_{ac}$, then it  is straightforward to
finally obtain that $w_{\pm}\simeq w \pm \lambda$, where $\lambda \ll w $.
Secondly, the position of the guiding center of the hybrid driven-LS mode is
now\cite{torma}
$X(t)=X_{0}+A \sin w_{+}t+B \sin w_{-}t$.\\
With  similar algebra as before, we introduce the damping
that the hybrid driven-LS mode undergoes due to scattering with the lattice ions.
The obtained expression for $w_{\pm}$ and $X(t)$ are now:
\begin{equation}
2w_{\pm}^{2}=(w^{2}_{1}+w_{2}^{2})\pm \sqrt{(w^{2}_{1}-w_{2}^{2})+16 \lambda^{2}w_{2}w_{1}}
\end{equation}
\begin{equation}
X(t)=X_{0}+e^{-\frac{\gamma}{2}t}[A \sin w_{+}t+B \sin w_{-}t]
\end{equation}
where $w^{2}_{1}=w^{2}-\frac{\gamma2}{4}\simeq w^{2}$
and $w^{2}_{2}=w_{ac}^{2}-\frac{\gamma2}{4}\simeq w_{ac}^{2}$,
considering that  $\gamma^{2} \ll w^{2}, w_{ac}^{2}$.
With the new obtained expression for $X(t)$ (Eq. 3) and according to the radiation-driven electron orbit
model, we obtain for the average distance advanced by the electron in a
scattering event,
\begin{eqnarray}
\Delta X(t)&=&\Delta X_{0}-e^{-\frac{\gamma}{2}\tau}A[\sin w_{+}\tau+ \sin w_{-}\tau]\nonumber\\
           &=&\Delta X_{0}-2A e^{-\frac{\gamma}{2}\tau}\sin w\tau \cos \lambda\tau
\end{eqnarray}
where we have considered that the amplitudes of both modes are similar, i.e.,
$A \simeq B$ and that $w_{\pm}\simeq w \pm \lambda$. The time, $\tau$,  according to the
radiation driven electron orbit  model\cite{drivenLS,inascirep},  is the "flight time", the time it takes the electron to jump due
to scattering from one orbit to another and  its value is given by $\tau=\frac{2 \pi}{w_{c}}$.
Following the same RIRO model and using  the obtained $\Delta X(t)$,
we end up with an expression for $R_{xx}$\cite{drivenLS,inascirep}:
\begin{equation}
R_{xx}  \propto \Delta X_{0}-2A e^{-\pi\frac{\gamma}{w_{c}}} \sin \left(2 \pi \frac{w}{w_{c}}\right)  \cos \left(2 \pi\frac{\lambda}{w_{c}}\right)
\end{equation}
where we want to stand out the essential part that explains the
appearance of the beating pattern in RIRO.
\begin{figure}
\centering \epsfxsize=2.8in \epsfysize=3.in
\epsffile{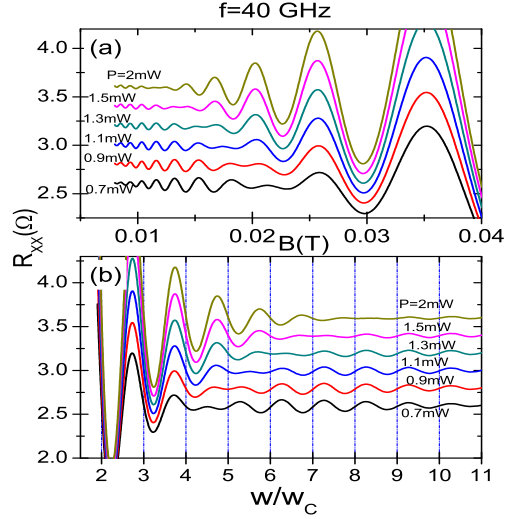}
 \caption{Same as in Fig. 3 but in function of radiation power. }
\end{figure}

In Fig. 1 we exhibit calculated $R_{xx}$ for a radiation frequency of 105 GHz. In
Fig. 1a, $R_{xx}$ vs $B$ and in Fig. 1b, $R_{xx}$ vs $w/w_{c}$. Apart from the usual RIRO we observe a beat at very low $B$
with a node around $B=0.02T$ in the upper panel. In the lower panel the vertical lines for integer values of
the abscissa indicate a phase change of $\pi$ in the $R_{xx} $ oscillations when
crossing the node. Eq. 5 readily explains the rise of a beat when $w_{+}$ is just slightly different from $w_{-}$
or in other words, when $\lambda \ll w$;
the change of phase in $\pi$ is due to the modulation of the slower function, i.e.,
the cosine function. In Fig. 2 we present the dependence of the  beating pattern
on the frequency, ranging from 30 GHz to 105 GHz. We observe a constant $B$-position
for the node irrespective of the frequency. We find again the explanation in Eq. 5. We can
tell that the node position depends on the cosine function where  $w$ does not show up. Thus,
the node position is immune to $w$. However any variation of the coupling constant $\lambda$
that shows up in the cosine will clearly affect the node position and even the number of
beats that can be observed.

\begin{figure}
\centering \epsfxsize=2.8in \epsfysize=3.5in
\epsffile{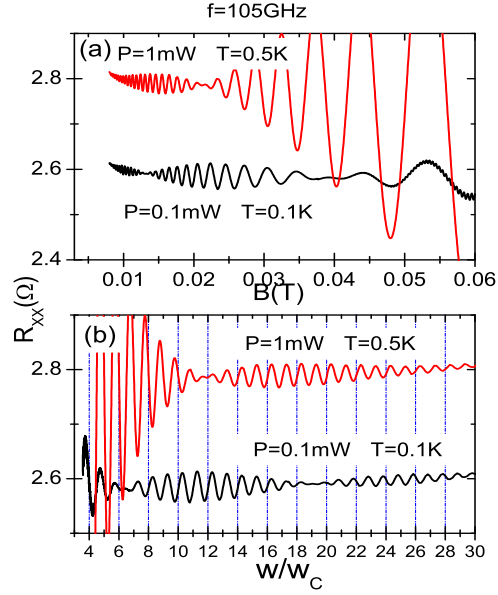}
 \caption{Dependence of the beating pattern on radiation power and temperaature for a radiation frequency of 105 GHz. In the upper panel
we exhibit irradiated $R_{xx}$ vs $B$ and the lower one $R_{xx}$ vs $w/w_{c}$. As expected, we observe  an extra beat and node when
going from $P=1mW$ and $T=0.5K$ to $P=0.1mW$ and $T=0.1K$. }
\end{figure}
In Fig. 3 we present the calculated results of the dependence of the beating pattern on temperature ($T$)
for a radiation frequency of 40 GHz and $T$ from 0.5K to 1.0K.
In Fig. 3a, we exhibit $R_{xx}$ vs $B$ and in Fig. 3b $R_{xx}$ vs $w/w_{c}$. For both panels the curves are shifted for
clarity. Interestingly enough,  as exhibited in the lower panel, the node position is not constant
and moves to lower $w/w_{c}$ for decreasing $T$, and in the same way the beat gets more intense. In the upper
panel we observe the opposite trend, the node moves to higher $B$ for decreasing $T$ but the intensity increase keeps
the same as in the lower one.  The displacement
of the node and the intensity variation indicate that a changing
temperature affects the driven-LS-phonon coupling and, in turn, $\lambda$. In Fig. 4 we study the beating pattern
in function of the radiation power ($P$). $P$ runs from 0.7 mW to 2 mW. Similarly
as in Fig.3, in panel a) we exhibit irradiated $R_{xx}$ vs $B$ and in panel b) the same
vs $w/w_{c}$. As with $T$, we observe a similar node displacement and beat intensity variation for decreasing $P$,
proving that $\lambda$ depends on $P$ too. Thus, both quantities, $T$ and $P$,
affect the beat in the same way: a decrease of any of them makes  the coupling stronger and $\lambda$ bigger.
And, on the other hand, an increase gives rise to a progressive destruction of the beat. The physical explanation
can be readily obtained from our model. A higher $T$  triggers a more intense scattering
between the electrons in their orbits and the lattice ions.  In the case of an increasing $P$, the
amplitude of the driven-LS oscillations gets also bigger, and in turn, the probability
for the electrons
in their orbits to be scattered  is also higher too. Thus, for both increasing quantities we
obtain a similar damping effect on
the driven-LS-acoustic phonon coupling that gets progressively destroyed. To study the dependence of $\lambda$, i.e., of the beating pattern  on $T$ and $P$,
we have developed
a phenomenological equation consisting.
  in adding a first order (linear) correction to $\lambda$ in the variation  of $T$ and $P$.
Thus, for $T$, $\lambda=\lambda_{0}-\lambda_{1}(T-T_{0})$ where $\lambda_{0}$ and  $\lambda_{1}$ are constants
and $T_{0}=0.5K$  in agreement with experimental values\cite{shi2}. And a similar equation for $P$, $\lambda=\lambda_{0}-\lambda_{1}(P-P_{0})$ and
 $P_{0}=0.7mW$.
The calculated results  are in qualitatively good agreement with experiments\cite{shi2}.

In Fig. 5 we exhibit the calculated results when  simultaneously changing $T$ and $P$. According to our
model, for instance in the case of decreasing,  this will lead to a strengthening of the beating pattern effect. Thus,  we would
expect the rise of more nodes and beats. This is presented  in Fig. 5 for a radiation frequency of 105 GHz; in the upper panel
$R_{xx}$ vs $B$ and the lower one $R_{xx}$ vs $w/w_{c}$. As expected, we observe the appearance of an extra beat and node when
going from $P=1mW$ and $T=0.5K$ to $P=0.1mW$ and $T=0.1K$. We observe the
jump of $\pi$ when crossing a node. Thus, in this figure we observe a total phase change of $2\pi$ when
crossing two nodes in a row. Again this is explained by the effect of modulation of the cosine function
on RIRO.

This work is supported by the MINECO (Spain) under grant
MAT2017-86717-P  and ITN Grant 234970 (EU).
Grupo de Matematicas aplicadas a la Materia Condensada, (UC3M),
Unidad Asociada al CSIC.

\end{document}